# Realizing Scalable Chemical Vapour Deposition of Monolayer Graphene Films on Iron with Concurrent Surface Hardening by *in situ* Observations

Bernhard Fickl,[1] Werner Artner,[2] Daniel Matulka,[1,2] Jakob Rath,[1,3] Martin Nastran,[1] Markus Hofer,[1] Raoul Blume,[4] Michael Hävecker,[4,5] Alexander Kirnbauer,[6] Florian Fahrnberger,[7] Herbert Hutter,[7] Dengsong Zhang,[8,*] Paul H. Mayrhofer,[6] Axel Knop-Gericke,[4,5] Beatriz Roldan Cuenya,[5] Robert Schlögl,[4,5] Christian Dipolt,[9] Dominik Eder,[1,*] Bernhard C. Bayer[1,*]

[1]Institute of Materials Chemistry, Technische Universität Wien (TU Wien), Getreidemarkt 9, 1060 Vienna, Austria

[2]X-Ray Center, Technische Universität Wien (TU Wien), Vienna, Austria

[3]Analytical Instrumentation Center (AIC), Technische Universität Wien (TU Wien), Vienna, Austria

[4]Max-Planck-Institut für Chemische Energiekonversion, Postfach 101365, Mülheim an der Ruhr 45413, Germany

[5]Department of Inorganic Chemistry, Fritz-Haber-Institut der Max-Planck Gesellschaft, Faradayweg 4-6, 14195 Berlin, Germany

[6]Institute of Materials Science and Technology, Technische Universität Wien (TU Wien), Vienna, Austria

[7]Institute of Chemical Technologies and Analytics, Technische Universität Wien (TU Wien), Vienna, Austria

[8]International Joint Laboratory of Catalytic Chemistry, State Key Laboratory of Advanced Special Steel, Innovation Institute of Carbon Neutrality, Research Center of Nanoscience and Technology, Department of Chemistry, College of Sciences, Shanghai University, Shanghai 200444, China

[9]Rübig GmbH & Co KG, Schafwiesenstraße 56, 4600 Wels, Austria

*Corresponding Authors: bernhard.bayer-skoff@tuwien.ac.at (Bernhard C. Bayer), dominik.eder@tuwien.ac.at (Dominik Eder), dszhang@shu.edu.cn (Dengsong Zhang)

**Abstract**

Graphene has been suggested as an ultimately thin functional coating for metallurgical alloys such as steels. However, even on pure iron (Fe), the parent phase of steels, growth of high quality graphene films remains largely elusive to date. We here report scalable chemical vapour deposition (CVD) of high quality monolayer graphene films on Fe substrates. To achieve this, we here elucidate the mechanisms of graphene growth on Fe using complementary *in situ* X-ray diffractometry (XRD) and *in situ* near ambient pressure X-ray photoelectron spectroscopy (NAP XPS) *during* our scalable CVD conditions. As key factors that set Fe apart from other common graphene CVD catalyst supports such as Ni or Cu, we identify that for Fe (i) carbothermal reduction of persistent Fe-oxides and (ii) kinetic balancing of carbon uptake into the Fe during CVD near the Fe-C eutectoid because of the complex multi-phased Fe-C phase diagram are critical. Additionally, we establish that the carbon uptake into the Fe during graphene CVD is not only important in terms of growth mechanism but can also be advantageously utilized for concurrent surface hardening of the Fe during the graphene CVD process akin to carburization/case hardening. Our work thereby forms a framework for controlled and scalable high-quality monolayer graphene film CVD on Fe incl. the introduction of concurrent surface hardening during graphene CVD.

## Introduction

Two-dimensional (2D) materials, incl. graphene and 2D hexagonal boron nitride, have been heralded as ultimately thin functional corrosion barrier coatings for modern metallurgical alloys, incl. steels.[1–4] This is because 2D materials can highly selectively impede transport of matter but enable transport of energy/charge between their substrate and their environment over ultimately small thickness scales of just one or a few atoms. For instance, graphene on steel could (due to graphene's record impermeability to corrosive species[5,6]) impede corrosive diffusional processes between the steel and its environment while (due to its high electrical conductivity[7,8]) still enabling highly efficient charge transfer between the steel and its environment to allow for, e.g., efficient current collector/electrode functionality with ultimately minimal coating thickness.-Such complementary barrier functionality is much harder to achieve with conventional, typically much thicker (>100 nm) barrier coatings.[9] Likewise, graphene coatings may offer additional functionality such as controlled wetting, anti-icing or biocompatibility.[3,10–13] Thus, substantial work has gone into coating metallurgical alloys and in particular steels with graphene as ultimately thin, functional barriers.[1–3] Target for such coatings is to produce graphene films on steels with complete coverage, controlled layer numbers and good interfacing to the steel substrate.[1–3] To date, however, only structurally imperfect graphene coatings with incomplete coverage, high defect levels, low control over layer numbers and incomplete interfacing to the substrate have been obtained on steels, be it from top-down liquid phase exfoliation[14–20] or bottom-up chemical vapour deposition (CVD).[21–28] Importantly, even on pure iron (Fe), the parent phase for all steels, to date no monolayered graphene films with complete coverage have been reported, let alone under scalable conditions.[29–41] This lack of graphene growth on even simple, pure Fe is thereby a clear hindrance to further advancing graphene growth on more complex, multi-element, multi-phased steels.

Filling this critical gap, we here report scalable CVD of monolayered graphene films on Fe substrates. Importantly, our here reported CVD conditions are scalable and compatible with current gas phase surface hardening/carburisation processes as used in the metallurgical industry. Consequently, we also demonstrate that our graphene CVD process also leads to concurrent surface hardening of the Fe substrates via carbon uptake into the Fe sub-surface and bulk. To achieve this goal of monolayer graphene film CVD on Fe, we here also elucidate the mechanisms of graphene growth on Fe using complementary *in situ* X-ray diffractometry (XRD) and *in situ* near ambient pressure (NAP) X-ray photoelectron spectroscopy (XPS) *during* our scalable CVD conditions to understand the complex interplay of the Fe's surface, sub-surface and bulk with the gaseous hydrocarbon CVD precursors and residual trace gases under kinetically-controlled CVD process conditions. In particular, we find that the controlled growth of high-quality monolayer graphene on iron has been challenging not only because of the non-trivial iron-carbon (Fe-C) phase diagram but also because of the inhibition of graphene growth due to persistent Fe surface oxidation. We investigate and overcome these challenges by our *in situ* characterisation-guided CVD process development. Our work thereby forms a holistic framework for process development of controlled and scalable high-quality monolayer graphene CVD on Fe-type substrates incl. introduction of concurrent surface hardening, which we expect to also lay the basis for subsequent, future expansion of graphene CVD coatings on persistently challenging steel substrates.

Generally, graphene CVD is a bottom-up approach in which gaseous precursors (mostly hydrocarbons) are flown at elevated temperatures (~400 °C to ~1000 °C) over the desired growth support, leading to precursor breakdown and then (under the right process conditions) graphene growth.[42–44] As prior work has shown,[45–47] unlike conventional CVD of classical µm-thick coatings where the substrate is comparatively "inert", in graphene CVD the growth substrate has a highly active catalytic role via surface catalytic activity and also bulk solubilities/diffusivities.[44,48] In particular substantial uptake of carbon, graphene's constituent element, into the growth substrate's bulk can occur during graphene CVD. This complicates graphene growth kinetics and requires close matching of CVD conditions (temperature profiles, precursor fluxes etc.) with the growth substrate. In the past, graphene CVD has been optimized for dedicated, often sacrificial high-purity Cu and Ni metal growth catalyst supports[45–47,49–51] to fully covering, layer-number-controlled, high quality graphene films. In comparison, graphene CVD on Fe has been significantly lagging behind.[29–41]

A first factor that sets apart Fe from other catalyst substrates is the more complex, multi-phased Fe-C phase diagram (Figure 1). As shown prior for Ni and Cu catalyst substrates based on *in situ* investigations, graphene CVD follows a bulk-mediated surface growth mechanism.[45–47] This means that graphene CVD is governed both by surface processes (gaseous precursor breakdown and reorganisation of surface species into graphene nuclei/domains) and bulk-mediation in which precursor supersaturation by diffusion on the surface, into the subsurface and, depending on kinetics, also into the bulk of the support must be reached before graphene nucleation/growth can occur.[44–48] Then, growth can proceed isothermally on the surface and/or via precipitation from the bulk upon cooling.[45–47] Importantly, for a given catalyst support with given C solubility the exact pathways of graphene CVD within the interplay of surface processes and bulk mediation can be kinetically controlled.[44,48] Key hereby is controlling the balance between incoming precursor flux, flux to the graphene's growth front and what flux is diffusing into the catalyst support bulk. Hereby, isothermal surface growth typically leads to better control over 2D materials layer numbers, quality and coverage, while precipitation upon cooling typically leads to undefined growth with inhomogeneous layer numbers and coverage and poorer crystalline quality when using gaseous precursors and standard CVD methods.[50,52] Compared to prior work on Ni and Cu catalyst supports, which were shown to remain single-phased during the entire graphene CVD process,[45–47] we find in this work that the here investigated Fe catalyst substrate can undergo substantial, temperature- and process-stage dependent phase transitions, e.g. body-centred-cubic (bcc) to face-centred-cubic (fcc) Fe during carbon feeding with strong increases in C solubility for >727 °C.[53,54] Such high, temperature-dependent solubility often favours precipitation from bulk 2D growth and thus typically yields 2D films of low quality when Fe-based supports are used.[50,55] As we demonstrate in this report, the key to overcome this limitation is the identification of kinetic conditions for CVD in terms of temperatures, precursor concentrations, and diffusion fluxes that facilitate predominantly isothermal growth on Fe nevertheless.

A second aspect that sets apart Fe catalysts from widely used Ni and Cu graphene growth catalyst supports, is Fe's propensity to readily form persistent surface oxides during Fe storage in ambient conditions before CVD and also from residual oxidizing species (trace oxygen and water) *in situ* during CVD. Surface oxides, however, typically suppress Fe catalytic ability for graphene CVD entirely or, at best, only lead to defective graphene.[56] Therefore, a reduction step with reductive gases, e.g. annealing in $H_2$, is typically employed before hydrocarbon exposure in graphene CVD and a reductive gas is also typically added during the hydrocarbon exposure to suppress *in situ* oxidation. In comparison to Ni and Cu support, we show here however that typical reduction conditions with $H_2$ are insufficient for Fe reduction under scalable CVD conditions but that also carbothermal reduction of Fe-oxides from hydrocarbon exposure is a key element in monolayer graphene CVD on Fe. This highlights that not only the kinetics for graphene growth (as eluded to above) but also the kinetics for reduction of persistent Fe-oxides must be controlled for graphene CVD development on Fe.

A third aspect that is of particular usefulness for Fe is that the carbon uptake into the sub-surface and bulk during graphene CVD (that we here also evidence using *in situ* XRD and XPS) is reminiscent of industrially widely applied carburization hardening (case hardening) for Fe/steels.[57] Exploring this aspect, we finally also demonstrate that under our optimized graphene CVD growth conditions, the remaining significant carbon uptake into the Fe bulk also leads to concurrent surface hardening of the Fe substrates. Thus, from a metallurgical application perspective, a beneficial interplay of concurrent graphene CVD and surface hardening is demonstrated here.

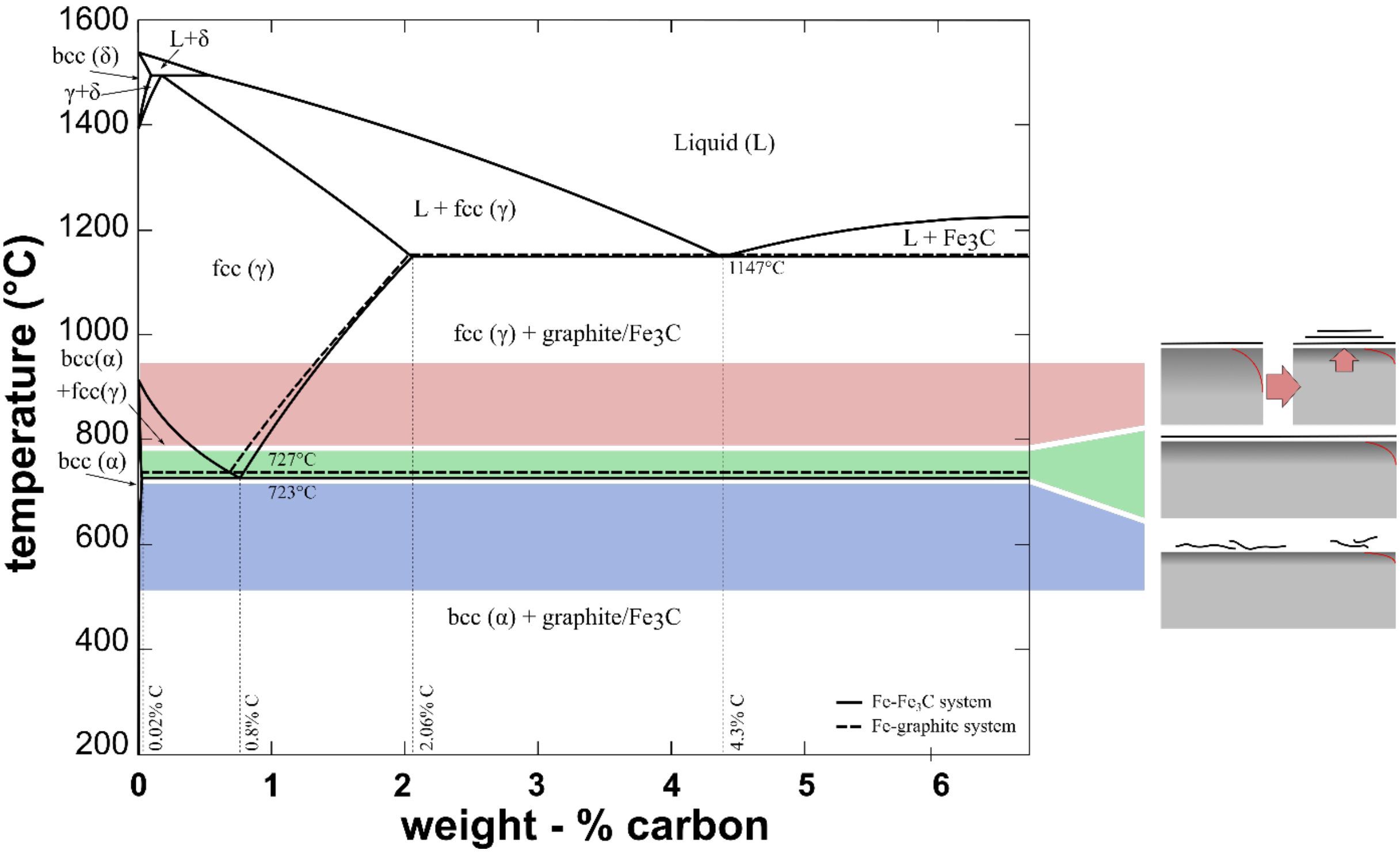

**Figure 1.** Schematic illustration of the Fe-C phase diagram for both Fe-graphite (dashed) and (metastable) Fe-$Fe_3C$ systems (adapted from OpenCalphad[58] and modified), with a schematic illustration of the main findings of this study regarding the interplay of CVD conditions and graphene growth results superimposed (see Discussion section). The metastable intermetallic $Fe_3C$ phase is at 6.67 weight-% C (at the right end of this diagram).

## Results

**Rationally designed CVD conditions.** To ensure fine control over the carbon flux we base our CVD recipe on $C_2H_2$ as the hydrocarbon source. The investigated process parameters are initially based on prior developed CVD conditions for Ni catalyst supports.[45,46] $C_2H_2$ has the advantage of dissociating readily and being active for graphene growth already at lower temperatures from ~450 °C.[45] Thus $C_2H_2$ can be employed at low and well controllable fluxes for graphene CVD. We here employ the $C_2H_2$ in a simple custom-built hot-wall quartz tube furnace with mass-flow-controlled $C_2H_2$ in-flux under medium-pressure CVD conditions obtained by a simple pump setup (base pressure $3\times10^{-3}$ mbar). As reductive gas we employ $H_2$ for pre-CVD reduction and also during hydrocarbon exposure. In the CVD process, samples are heated in ~1 mbar $H_2$ (250 sccm flow) to their target temperatures of 500 °C to 800 °C at ~100 °C/min heating rate and, upon reaching the desired temperature, undergo an annealing step in 1 mbar $H_2$ (250 sccm flow) for 30 min. Then 0.1 to 10 sccm $C_2H_2$ are added to the $H_2$ (250 sccm) for the growth step for another 30 min. Subsequently, $C_2H_2$ and the heater are switched off simultaneously, and samples are left to cool naturally in ~1 mbar $H_2$ (natural cooling at ~35 °C/min to ~300 °C, then ~15 min to room temperature; split tube furnace around quartz tube opened during cooling). We emphasize that such CVD conditions are directly compatible with common carburisation hardening conditions in industrial surface hardening processes[57] and are thus intrinsically industrially scalable. As Fe samples we employ high purity 100 µm thick Fe foils (Alfa Aesar Puratonic®, 99.995%). We deliberately chose the comparatively high thickness of the Fe foils to also account for bulk effects that have been shown to play an important role for Ni catalysts.[44,48] For further information on experimental details see Methods Section.

**Optimization of Graphene CVD Results.** We first describe a survey of CVD parameter space to illustrate our optimized growth results before providing experimental (*in situ*) insights into the corresponding growth mechanisms further below. Figure 2 shows optical microscopy images (left) and corresponding, spot-localized Raman spectra (right, spot localisation indicated by coloured spectra/spots) of growth results on the Fe supports from the above describe CVD conditions for an intermediate $C_2H_2$ flux of 1 sccm as a function of growth temperature from 500 °C to 800 °C (and referenced against as received Fe foil). For the as-received Fe foil, we find in optical microscopy and Raman[59,60] (green trace) that the foils have formed surface Fe-oxides from storage in ambient air. After 500 °C CVD we find the Fe foil to be inhomogeneously covered by nanocrystalline graphite (red trace: intensity ratio D/G >2 and very low 2D intensity[61,62]) and amorphous carbon (blue trace: merged D and G, no 2D[63]) regions. Under these nanocrystalline graphene and amorphous carbon regions, no signs of remaining Fe-oxide are detected in Raman, implying localized reduction of the Fe-oxides during the CVD process. The graphitisation level of the carbon deposits from the 500 °C growth temperature indicates insufficient thermal activation for healing out defects in the growing carbon film.[64] With increasing temperature to 600 °C, we accordingly find an improvement in graphitisation levels: We grow inhomogeneous multilayer graphene films at 600 °C without (blue trace: intensity ratio D/G ~0.3; 2D/G ~0.7)[65] and with remaining Fe-oxides (red trace) and small graphene-bare Fe-oxide regions (green trace). When further increasing the growth temperature to 700 °C we find further improvements in graphitization, indicated by a further reduction in D/G ratio to <0.2.[65] Additionally, we now find an inhomogeneous mixture of multilayer graphene (blue trace) as well as monolayer graphene regions (red trace: 2D/G ~1.5).[65] Notably however, persistent Fe-oxides are still detected (green trace). Further increasing the growth temperature to 750 °C we find clear improvements in homogeneity, importantly towards predominantly monolayer graphene growth of high quality (red trace: D/G <0.05; 2D/G ~2).[65] Quantitatively, we estimate monolayer graphene sample coverage to ~70% (based on optical micrographs and Raman analysis). The remaining non-monolayer-graphene areas are comprised of isolated multilayer graphene islands to ~10% sample coverage (dark spots in leftmost optical micrograph, blue trace) and remaining Fe-oxide regions (~20%), which are however void of graphene or carbon coverage (green trace). Thereby, monolayer and multilayer graphene regions combined have a coverage of ~80% on the iron. We further confirm our monolayer assignment of these graphene films (and exclude formation of turbostratic graphite) via a standard polymer-assisted transfer[66] of the films from their Fe support onto 90 nm $SiO_2$-coated Si wafers and further Raman and optical microscopy data in Supporting Figures S1 and S2. Interestingly, when further increasing the CVD temperature to 800 °C, we do *not* observe further improvements in controlled graphene coverage but instead obtain comparatively much more inhomogeneous carbon films with only a small fraction of monolayer graphene coverage (red trace) but large fractions of multilayer graphene growth (dark patches in leftmost image, blue trace) as well as bare remaining Fe-oxide regions (green trace). Notably however, growth at 800 °C retains similarly high graphitic quality[65] (D/G <0.05) as for 750 °C. This indicates that for the 800 °C not graphitisation but other growth mechanistic factors prevent predominant monolayer graphene film growth.

Taking the so far best monolayer graphene results from CVD at 750 °C at 1 sccm $C_2H_2$ from Figure 2 as optimized reference point, we then compare the effect of $C_2H_2$ flux in Supporting

Figure S3. We however find that growth at a lower $C_2H_2$ flux of ~0.1 sccm leads to only monolayer island growth with large areas of the substrate left covered in Fe-oxide. This indicates insufficient carbon flux. Conversely, growth at increased 10 sccm $C_2H_2$ flux leads to a relative increase in large area multilayer graphitic growth, thus implying that 10 sccm $C_2H_2$ represent a too high carbon flux for predominant monolayer growth. This suggests that overall at 750 °C the 1 sccm $C_2H_2$ flux represents, under the screened conditions, an optimized balance of incoming precursor carbon flux, carbon flux to the graphene's growth front and what carbon flux is diffused into catalyst support bulk toward best monolayer graphene growth results.[44]

We note that overall our optimized results here at 750 °C in Figure 2 go beyond prior graphene CVD on Fe in terms of quality and monolayer coverage, particularly under scalable CVD conditions.[29–41]

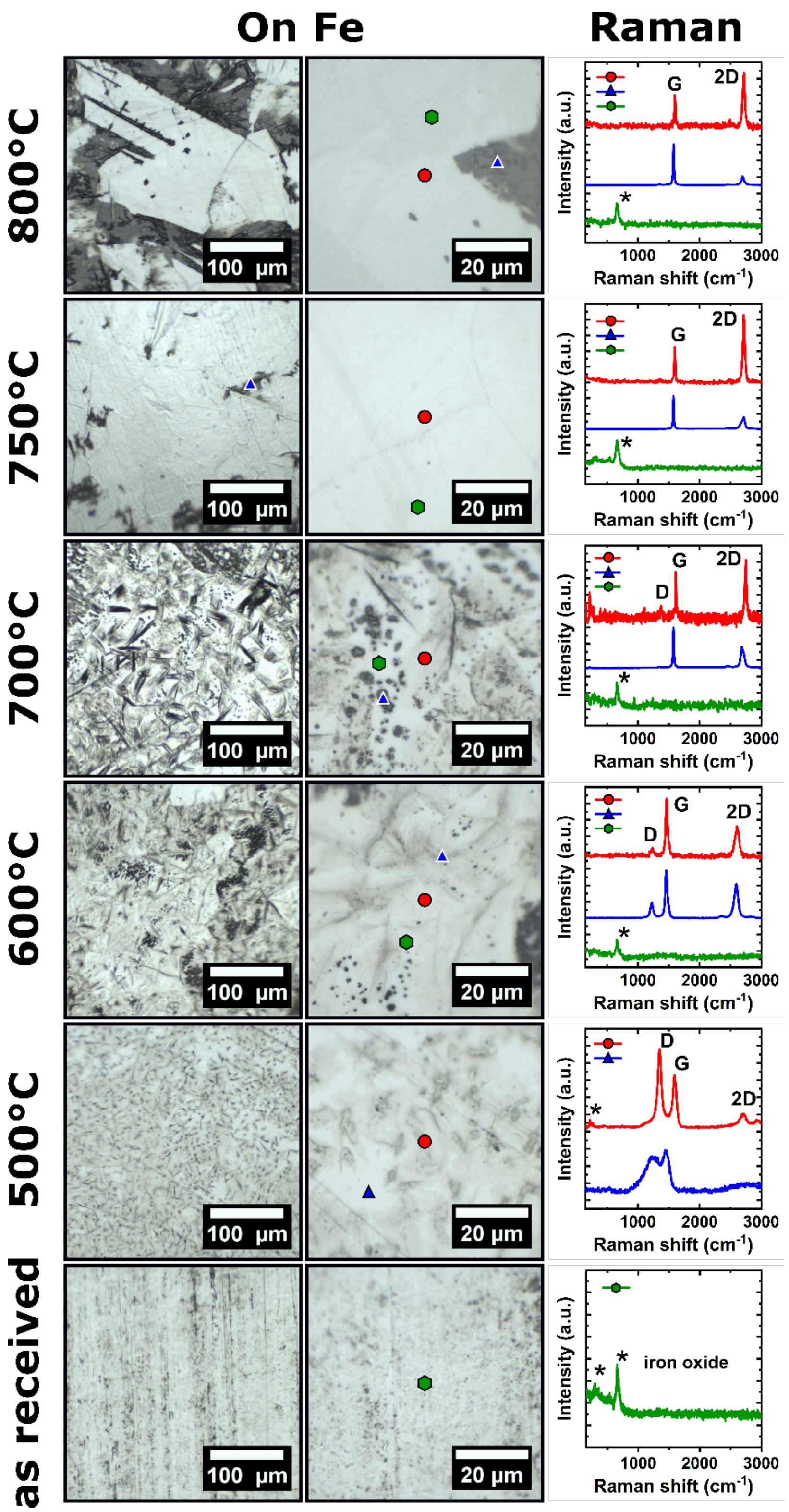

**Figure 2.** Optical micrographs (left and middle panels) at different magnifications and spot-localized Raman spectra (right panel, spot localisation indicated by coloured spectra/spots) of as received 100 µm Fe foils and growth results on Fe after CVD with 1 sccm $C_2H_2$ at temperatures from 500 °C to 800 °C. Carbon related D, G and 2D Raman peaks[65] are labelled and iron oxide related peaks[59,60] are indicated by "*" in the Raman spectra.

**Investigation of Growth Mechanisms.** After having established our optimized graphene CVD protocol on Fe, we now turn to elucidating the underlying mechanisms including *in situ* investigations. We first investigate the key importance of, as we find, carbothermal surface Fe-oxide reduction during CVD growth, before investigating the Fe-C phase and surface chemistry evolution in/on the Fe catalyst support foils during our optimized CVD conditions by complementary *in situ* XRD and *in situ* NAP XPS.

**Importance of Carbothermal Reduction of Fe-oxides.** Our data in Figure 2 indicates that the presence of persistent surface Fe-oxides, which we detect as a minority surface coverage under practically all CVD conditions tested, is a remaining unfavourable factor in our graphene growth on Fe. Such Fe-oxides are detrimental to graphene growth because generally oxides are known to be much less suited to catalyse high-quality graphene during CVD.[67,68] Surface Fe-oxides can either form from ambient air during Fe catalyst support storage prior to CVD incl. subsequent Fe-oxide crystallisation during the high temperature CVD process and/or from *in situ* oxidation of the Fe catalyst support from residual trace gases such as $O_2$ or water during the CVD process.[69] To counter both processes and reduce such Fe-surface oxides most CVD recipes, including ours, use a dedicated reductive pre-treatment step (here 1 mbar $H_2$) and/or a reductive ad-gas (here 1 mbar $H_2$) being present throughout the entire CVD process. Additionally, however, carbothermal reduction of the Fe-oxides from the hydrocarbon source (here $C_2H_2$) may also occur, but is however commonly not explicitly considered. Compared to other established graphene catalysts, Fe-oxides are however known to be more stable and intrinsically harder to reduce than in comparison under their respective CVD conditions readily reducible Ni-oxides[45,46] and Cu-oxides.[47]

To therefore disentangle Fe-oxide formation and reduction processes under our CVD conditions, we conduct cross-check experiments: In Supporting Figure S4 we present optical microscopy and Raman spectroscopy results for 100 μm Fe foils that underwent the CVD process at 750 °C but *without* $C_2H_2$ addition i.e. samples only underwent annealing in $H_2$. For these samples we find no graphene growth (as expected due to no $C_2H_2$ exposure), however, despite the reducing $H_2$ conditions, the presence of a surface Fe-oxide over the entire Fe foil surface is detected. Together with the observed presence of an initial surface Fe-oxide in our as received foils (Figure 2), this implies that under our conditions (and in our CVD furnace) the $H_2$ alone is not sufficient for initially present Fe-oxide reduction and suggests that the $C_2H_2$ under our process conditions has not only the role of graphene growth precursor but also of a carbothermal reduction agent, since only with $C_2H_2$ introduction as in Figure 2 the majority of the Fe has been reduced (as indirectly evidenced by the observed carbon growth). Our results in Figure 2 nevertheless show that through carbothermal reduction good graphene films can already be achieved by simple and well scalable medium vacuum conditions (~$10^{-3}$ mbar base pressure), as employed here, also for Fe.

**Fe-C Phase Dynamics during CVD by *in situ* XRD.** After having *ex situ* investigated the importance of enabling surface Fe-oxide reduction incl. carbothermal reduction, we now turn to investigate the Fe-C phase dynamics during graphene CVD on Fe catalyst supports. Supporting Figure S5 shows *ex situ* XRD patterns of the Fe supports before and after CVD processing corresponding to Figure 2. As received foils are at room-temperature of phase-pure metallic body-centered-cubic (bcc) Fe (α-Fe) structure in accordance with the phase diagram (Figure 1). No Fe-oxides are detected in XRD, further confirming that the Fe-oxides in Raman (Figure 2) are surface oxides. After CVD and subsequent cooling to room temperature, we find for all growth temperatures the majority phase to be bcc-Fe but a minority Fe-carbide $Fe_3C$ phase has been formed additionally during CVD (Note the square-root intensity scale in Supporting Figure S5 that strongly emphasizes this minority $Fe_3C$ phase. Rietveld refinement puts a maximum phase contribution of $Fe_3C$ to ~12 %.). A graphite-related (002) peak is detected as function of growth temperature in accordance with the presence and roughly the amount of multilayer graphene compared to Figure 2. The observation of a Fe-carbide signal in Supporting Figure S5 implies that during the CVD process the Fe catalyst support is subjected to an influx of carbon into the catalyst bulk, resulting in the observed formation of an additional $Fe_3C$ phase. To investigate this phase evolution further, we therefore turn in Figure 3 to process-step-resolved *in situ* XRD measurements during our optimized CVD conditions at ~750 °C to reveal the phase evolution of the Fe catalyst support *during* each CVD process step. Here we find that the initial bcc Fe retains its bcc Fe structure during the $H_2$ annealing step at 750 °C but during the subsequent $C_2H_2$ exposure at 750 °C undergoes a phase transition to face-centered-cubic Fe (γ-Fe) phase. This is direct evidence for the carbon influx into the catalyst bulk during the $C_2H_2$ exposure, because according to the phase diagram (Figure 1) with increasing carbon concentration in the Fe a phase transition from bcc to fcc Fe occurs for growth temperatures above the eutectoid temperature of ~723 °C.[53,54] Concurrently we observe the emergence of a graphite peak during $C_2H_2$ exposure, giving direct evidence of isothermal graphene growth via our *in situ* XRD experiments. The observation of fcc Fe as the predominant phase during graphene CVD reaffirms that *ex situ* XRD measurements such as in Supporting Figure S5 can not necessarily capture the relevant phase evolution (as no fcc Fe has been detected in Supporting Figure S5 at all) but that *in situ* experiments are necessary.[53,54] After CVD and after cooling to room temperature we observe that the Fe has reverted back to bcc Fe (again in agreement with the phase diagram in Figure 1). During our *in situ* XRD measurements no indication for substantial $Fe_3C$ formation during the CVD process was observed. We note, however, that our *in situ* XRD runs in Figure 3 employed a Cu anode which for Fe samples results in higher background due to fluorescence, while our *ex situ* XRD data in Supporting Figure S5 was measured with a Cr anode that allows for higher sensitivity.[70] Conversely, when *ex situ* remeasuring our *in situ* sample from Figure 3 with a Cr anode after CVD we accordingly measure a very small $Fe_3C$ signal, which could have formed either during $C_2H_2$ exposure or during cooling (Top pattern in Figure 3. Note the square-root intensity scale in Figure 3. Rietveld refinement puts $Fe_3C$ to an upper limit of ~12 %, fully consistent with the *ex situ* growth in Supporting Figure S5). Combined, our *in situ* XRD data at optimized CVD conditions therefore indicates that fcc Fe is the majority phase in the Fe foils during growth and that a minority $Fe_3C$ phase could possibly also be present during growth. In either case, the above XRD data has confirmed carbon uptake into Fe as an important factor during growth (which results in the bcc to fcc Fe transition and $Fe_3C$ formation) and that graphene growth occurs (at least partially) isothermally. The time resolution of our *in situ* XRD measurements is, however, not sufficient to disentangle dynamics of the isothermal graphene growth and answer if growth via precipitation of prior dissolved carbon during cooling also contributes to graphene growth. To answer these questions we turn to *in situ* NAP XPS with better time resolution.

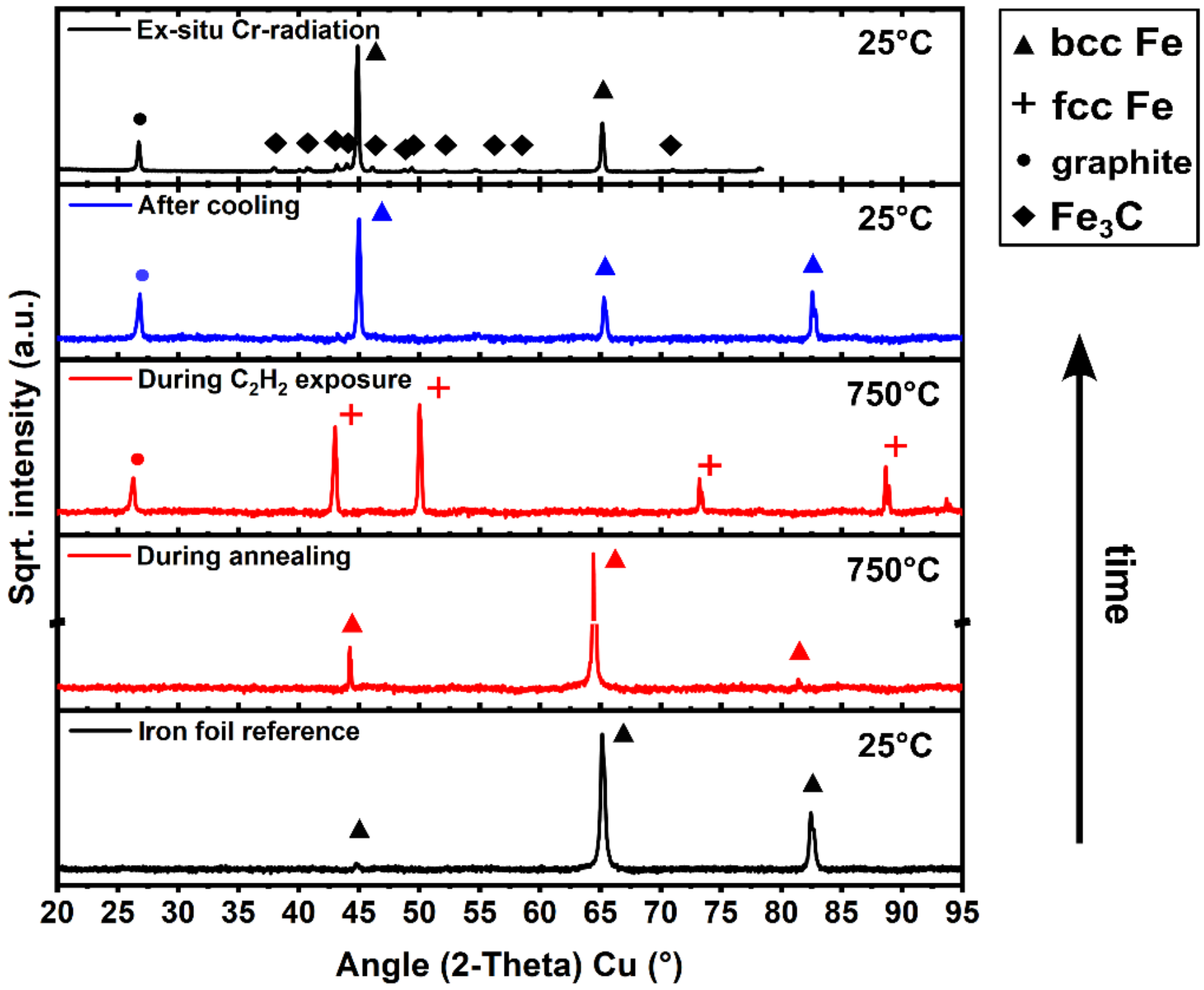

**Figure 3.** Process-step resolved *in situ* XRD patterns of Fe a catalyst supports during CVD at ~750 °C. Process step conditions (from bottom to top) are indicated. Salient phases identified are indicated. (International Centre for Diffraction Data (ICDD), PDF-5+ database, powder diffraction file entry: bcc-Fe-ambient 04-015-8438; bcc-Fe-non-ambient 040-17-5839; fcc-Fe-non-ambient 04-003-1443; Carbon/graphite 04-016-0554; $Fe_3C$ 04-007-0422) The *in situ* XRD patterns were measured with a Cu anode, resulting in higher background signal for Fe,[70] while the uppermost pattern was measured *ex situ* after CVD with a Cr anode thus also detecting a minority $Fe_3C$ phase that was below the sensitivity of the *in situ* Cu anode measurements (Cr anode pattern recalculated to 2-Theta angles comparable to Cu anode dataset). Note that the intensity scale is plotted in square-root and with intensity scale interruption(s) for better visualization of minority phase signals. Cr anode 2-Theta data was recalculated to Cu anode for better comparison.

**Surface evolution during CVD by *in situ* NAP XPS.** We employ *in situ* NAP XPS to study the surface evolution of carbon and Fe and their interactions throughout the graphene growth process at the same nominal condition as in our optimized growth from Figure 2. Notably, while we investigate the bulk of the Fe sample volume in our XRD measurements in Figure 3 above (and at only tens of minutes time resolution), with the XPS measurements we probe the uppermost few nm of the sample surface and sub-surface and at a time resolution of seconds.[51]

Figure 4a,b shows time-resolved C1s spectral evolution during $C_2H_2$ exposure step at 750 °C. The Fe sample is initially fully clean from adventitious carbon (removed during the $H_2$ pre-treatment) as evidenced by the flat C1st spectrum at 0 s in Figure 4a,b. Upon $C_2H_2$ exposure we first observe the emergence of a peak at 283.2 eV, starting at ~17 s. This 283.2 eV peak we ascribe to carbon bonded at iron surface sites based on previous work using Ni substrates.[46,71] This component has also an asymmetric shoulder towards higher binding energies at 283.7 eV (see in particular Figure 4b), which becomes more visible with time. This shoulder can be ascribed to an additional C1s component at 283.7 eV, which we attribute to carbon dissolved in Fe, again based on prior work.[46,71] Both the 283.2 eV component and 283.7 eV shoulder are thereby direct signs of carbon influx into the Fe, in excellent accordance with the XRD data above. We label both 283.2 eV and 283.7 eV components therefore as "Fe-C". Notably, both Fe-C components (283.2 eV, 283.7 eV) precede the first emergence of the C1s component of $sp^2$ graphene at 284.5 eV which emerges only after an incubation time after $C_2H_2$ introduction of ~51 s at 750 °C. Thereby the Fe-C 283.2 eV and 283.7 eV components are indicative of the necessary carbon influx into the Fe subsurface *before* graphene nucleation can occur. After first emergence of the $sp^2$ graphene at 284.5 eV signal at ~51 s, the graphene $sp^2$ signal overtakes the Fe-C components in intensity after ~113 s and then continues to rise with increasing $C_2H_2$ exposure time. This is further direct evidence of isothermal graphene growth on the Fe. (In Supporting Figure S6 we show detailed C1s components fits to the experimental XPS data.)

To resolve if carbon precipitation upon cooling also contributes to graphene growth from Fe under our optimized conditions, we also follow the C1s evolution after $C_2H_2$ shut off during the cooling step (~50 °C/min in $H_2$) in a time-resolved fashion in Figure 4c. We find only a small increase of graphene C1s signal at 284.5 eV during cooling (by ~18 %), showing thereby that under our growth conditions additional graphene formation by precipitation of prior dissolved carbon from Fe upon temperature cooling is limited. This links excellently with the observed only minor multilayer graphene coverage in Figure 2 at optimized monolayer graphene growth conditions at 750 °C.

The *in situ* NAP XPS data thereby indicates that the growth kinetics for our 750 °C growth on Fe are well controlled towards almost exclusive isothermal graphene growth with minimal additional graphene growth by precipitation upon cooling. In line with the XRD data however also for these kinetic conditions a significant carbon uptake into the Fe sub-surface (and from XRD Fe bulk) is evidenced as part of the graphene growth process. While this carbon reservoir is crucial to the isothermal graphene growth process at 750°C, it contributes only slightly to graphene growth via minor precipitation upon cooling.

A corollary to this finding is that when we increase the Fe "reservoir" for carbon uptake into the Fe with otherwise similar carbon feeding flux, we should change the growth kinetics toward a higher contribution of precipitation upon cooling growth. We test this hypothesis by measuring *in situ* NAP XPS also during higher temperature exposure at 800 °C. Based on the phase diagram at 800 °C (Figure 1) we would expect a higher carbon solubility in Fe and thus a larger free "reservoir" to accommodate carbon atoms in the Fe at the higher temperature. This should, for instance, directly translate to a longer filling period of this "reservoir" and thus a longer carbon uptake period before graphene nucleation. This is because the prefilling of the „reservoir" competes with reaching the surface supersaturation with carbon necessary for isothermal graphene nucleation/growth. Following this line of argument, we indeed find that the 800 °C growth temperature leads in the *in situ* C1s data during $C_2H_2$ exposure at 800 °C in Supporting Figure S7a to an increased incubation period of ~10 min (i.e. ~10-times longer compared to only ~51 s at 750 °C) during which only the Fe-C components (283.2 eV, 283.7 eV) are visible before the graphene $sp^2$ signal at 284.5 eV appears and graphene isothermally grows. Consistently, upon cooling from 800 °C after $C_2H_2$ exposure, a larger rise of the graphene C1s signal (284.5 eV) by 64 % is evidenced (compared to only 18 % at 750 °C), confirming a larger contribution of precipitation upon cooling to overall graphene growth at 800 °C. This is in excellent agreement with the increased multilayer fraction for growth results for 800 °C temperature compared to optimized 750 °C also in the *ex situ* data in Figure 2.

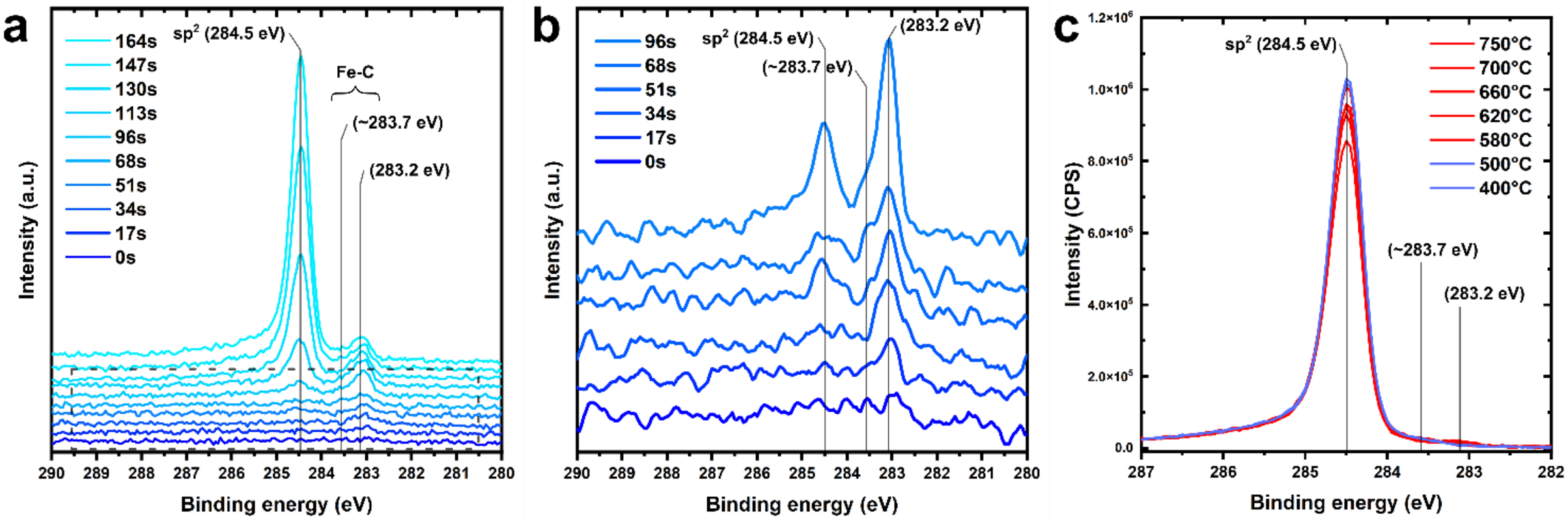

**Figure 4.** (a) C1s time-resolved *in situ* NAP XPS spectra during $C_2H_2$ exposure at 750 °C. Salient C1s components are indicated. (b) Zoom-in on corresponding region in panel (a). (c) C1s time-resolved after $C_2H_2$ shut off during cooling in $H_2$ from 750 °C.

**Concurrent Surface Hardening during Graphene CVD.** After having established the key role of carbon influx into the Fe subsurface and bulk during the graphene CVD process via our (*in situ*) observations above, we finally also probe the technological implication of this carbon influx. In particular, the here clearly observed carbon uptake into Fe is highly reminiscent of industrially widely applied carburization hardening (case hardening) processes for Fe/steels.[57] Comparing a Fe sample that underwent optimized graphene CVD at 750 °C with a sample that underwent similar $H_2$ annealing at 750 °C but without the $C_2H_2$ exposure step (i.e. no graphene growth), we show in Supporting Figure S8 using depth-profiling of the carbon signal via time-of-flight secondary-ion-mass-spectrometry (ToF-SIMS) that significant carbon uptake into the Fe bulk from the $C_2H_2$ exposure has taken place (at least to ~1 µm depth) compared to a practically carbon-free only $H_2$ annealed Fe reference sample. This *ex situ* data is thereby in excellent agreement with the (*in situ*) XRD and XPS data above. Consequently, we finally test which effect on Fe surface hardness this carbon uptake has: Employing nanoindentation measurements in Figure 5, we obtain hardness values for a graphene/Fe sample after optimized graphene CVD at 750 °C against an only $H_2$-annealed Fe reference sample. The data in Figure 5 clearly shows a drastic increase in hardness by ~130% that results from the carbon influx during graphene CVD. We thereby establish that surface hardening occurs *concurrent* to the graphene CVD process.

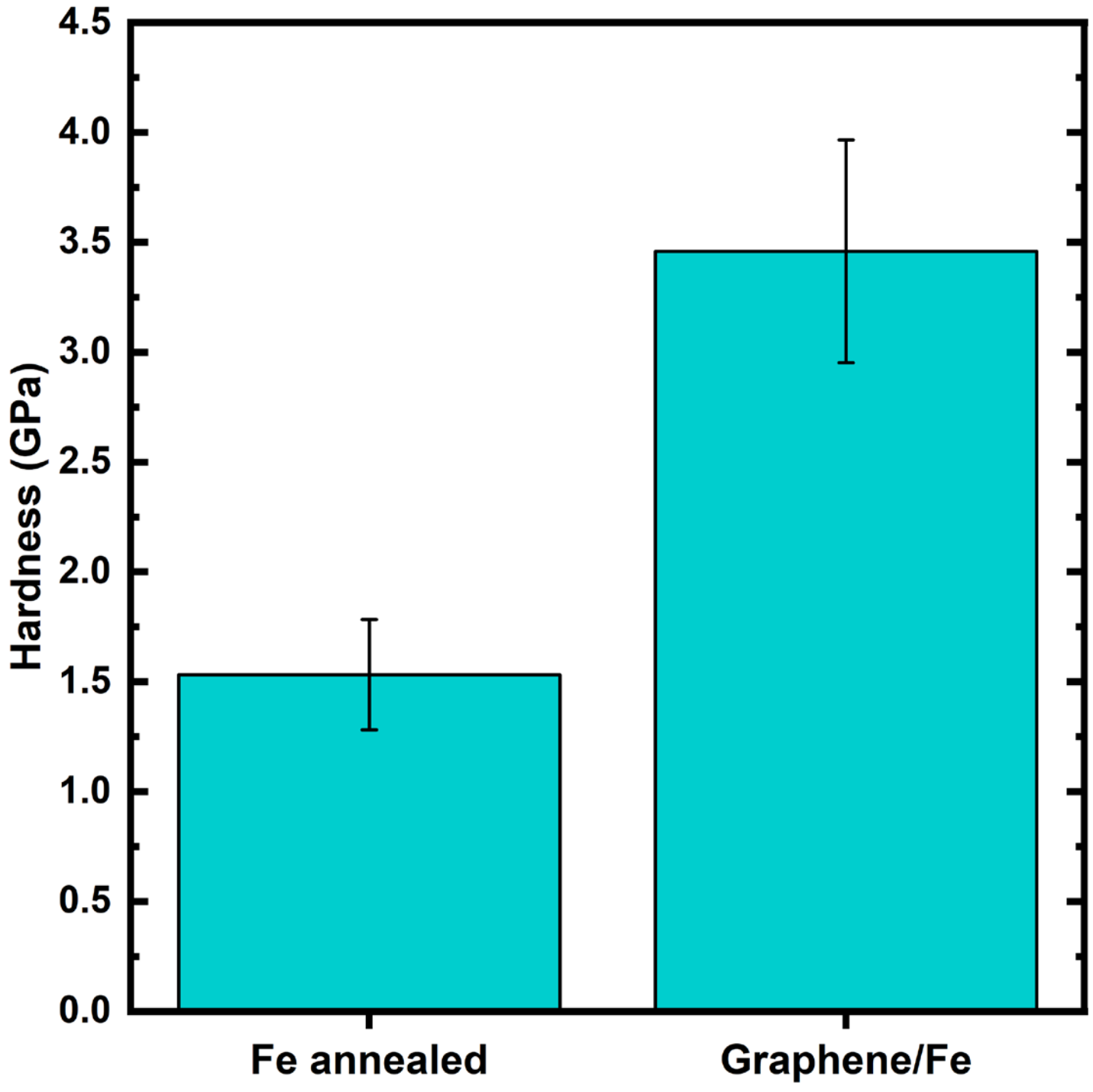

**Figure 5.** Hardness values from nanoindentation experiments for a graphene/Fe sample after optimized graphene CVD at 750 °C (1 sccm $C_2H_2$, Figure 2) against an only $H_2$-annealed Fe reference sample (i.e. without $C_2H_2$ exposure), elucidating a surface hardening effect concurrent to graphene growth under our optimized CVD conditions.

## Discussion

Our study here provides a general framework for optimizing graphene CVD on Fe. We illustrate our findings as schematic sketches to the Fe-C phase diagram in Figure 1.

Our data implies that at growth temperatures well below the Fe-C eutectoid (~723 °C), graphene growth on Fe is restricted by insufficient energy to nucleate and sustain high quality graphene growth, explaining the graphene growth of low structural quality in Figure 2 below 750 °C. At ~750 °C we find that near the 723 °C eutectoid a kinetic balance between well crystallized isothermal graphene growth with only minimal additional carbon precipitation upon cooling can be achieved, resulting in our optimized conditions for high quality graphene growth with unprecedented monolayer coverage on Fe. For higher temperatures (≥800 °C), while structural graphene quality may further improve from the additional thermal energy, the drastically increasing carbon uptake into the Fe support during graphene CVD results however in a hard to control increasing fraction of multilayer graphene growth from precipitation upon cooling. Thus, the higher growth temperatures (≥800 °C) again lead to worsened control over monolayer graphene coverage. Generally, finding a kinetic balance for graphene growth on Fe is critical for recipe development. We expect the above rationale to be applicable as a general guideline for the Fe-C system, albeit particular conditions will need adjustments for, e.g., sample sizes (i.e. more/less Fe volume to prefill during incubation time before graphene nucleation), hydrocarbon type (with $C_2H_2$ fairly reactive) and desired growth times. We expect the general rationale also to hold for steels, although the effect of the multiple add-elements in modern steels will require further consideration individually.

Another general finding pertaining to Fe is the here reported importance of controlling persistent Fe-surface oxides that can inhibit graphene growth and are harder to remove than in typical Ni- or Cu-based graphene recipes (which we can use without signs of persistent oxides on Ni or Cu in our CVD system[3,66]). We show however that, also aided by the here implicated carbothermal reduction during hydrocarbon exposure, also simple and cheap scalable medium vacuum conditions, as used in this study, are sufficient to account for this higher propensity of Fe-oxide formation compared to other typical graphene growth substrates.

Finally, we demonstrate that the carbon uptake during graphene CVD is not only relevant for a more complete mechanistic understanding of graphene CVD on Fe but also has, via the here introduced concurrent surface hardening during graphene CVD, technologically beneficial implications. In particular, we here demonstrate the potential of development of combined graphene growth and surface hardening processes for metallurgical materials, which is an aspect that remained little addressed in the literature.

## Conclusions

In summary, we have developed a CVD process for the growth of graphene on iron substrates, that can produce high quality monolayer graphene films with monolayer coverage of ~70% and total graphene coverage of ~80% under scalable CVD conditions. This represents a sizeable improvement of graphene CVD on Fe and is a prerequisite for growing graphene on more complex multi-element iron alloys such as steels. To obtain direct insights into the underlying growth mechanisms, we have followed the entire graphene CVD process on Fe using complementary *in situ* techniques to probe bulk crystallographic (*in situ* XRD) and surface chemical (*in situ* NAP XPS) evolution during CVD. Using this approach we identified that specifically (i) carbothermal reduction of persistent Fe-oxides and (ii) kinetic balancing of carbon uptake into the Fe during CVD near the Fe-C eutectoid are critical for high quality monolayer graphene CVD. Furthermore, we demonstrated that the carbon diffusion into the Fe is not only interesting from a growth mechanistic point of view but akin to industrial surface hardening processes (carburization/case hardening) and as such can be beneficially utilized for establishing concurrent graphene CVD and surface hardening processes, as we also here demonstrated.

## Methods

**Graphene CVD**. We employ a custom-built hot-wall CVD setup with a commercial split-tube furnace (Carbolite Gero Split tube furnace - HZS 12/600) around a quartz tube (GVB, EN08NB) for heating and temperature control and a combined, small turbomolecular pump/rotary vane pump stage (turbomolecular: VARIAN, Turbo-V 70LP, rotary: Vacuubrand RZ 2.5). In this configuration the base pressure of the CVD system is ~$3\times10^{-3}$ mbar. For a typical CVD run, samples undergo annealing in ~1 mbar $H_2$ (~250 sccm, Messer, 6.0 purity, 99.9999%) at the respective growth temperature (500°C to 800°C) for 30 min, after which 0.1 to 10 sccm of the carbon precursor acetylene ($C_2H_2$, Messer 2.6, 99.6% purity) is added for another 30 min. The samples are then left to cool naturally in ~1 mbar $H_2$ with the split-tube furnace heaters opened around the quartz tube. $H_2$ flow is controlled by a manual flow meter (Vögtlin Instruments GmbH, Q-Flow series) while $C_2H_2$ flow is controlled by a digital mass flow controller (MFC, Bronkhorst EL-flow select). We use 100 µm thick (Alfa Aesar Puratonic® 99.995%) polycrystalline iron foils as catalytic growth substrate.

***Ex situ* Characterization.** Samples are characterized *ex situ* via optical microscopy and Raman spectroscopy (WITec alpha 300 RSA+) after CVD. Laser wavelength 532 nm, laser power 10 mW, spot size ~2µm. *Ex situ* XRD measurements were conducted with a PANalytical X´Pert Pro multi-purpose diffractometer (MPD) with a standard rotating stage and chromium (Cr) anode as X-ray source with a wavelength of 2.26 Å. Presented Cr anode *ex situ* XRD patterns were scaled to make them comparable to the *in situ* XRD Cu anode datasets. While most characterisation investigated the graphene growth results directly on their Fe growth substrates, for selected samples graphene film transfer[66] was done using a polymethylmethacrylate/ethyl acetate mixture for drop casting a sacrificial polymer layer on top the graphene/Fe foil sample, followed by a bubbling transfer procedure,[72] before transferring the film onto a $SiO_2$(90nm)/Si substrate and dissolving the PMMA layer in acetone. For the bubbling transfer the PMMA coated graphene/Fe foil is dipped into 0.5 molar $K_2SO_4$ together with a glassy carbon electrode. A voltage of about 4-5 V is applied with the glassy carbon electrode acting as the anode and the graphene/Fe foil acting as the cathode. Hydrogen bubbles are formed between the iron foil and the PMMA supported graphene, separating the graphene from the substrate. Transferred graphene is characterized by Raman spectroscopy[65] and optical contrast analysis following a previously reported method.[73] Graphene coverage was calculated using visual measurements (thresholding of optical microscopy image) of graphene films on Fe and a transferred graphene film on a $SiO_2$(90nm)/Si substrate. For the transferred films, coverage is potentially underestimated due to damage of the film during transfer.

***In situ* XRD.** *In situ* XRD patterns were recorded on a PANalytical X´Pert Pro multi-purpose diffractometer (MPD) in Bragg Brentano geometry outfitted with an environmental cold-wall heating chamber (Anton Paar HTK 1200N) that can indirectly heat samples via a heating wire to up to ~1200°C and features atmospheric control through gas and vacuum regulation. Samples were placed on a ceramic sample holder and temperature was monitored via a RhPt thermo couple. The anode material used as X-ray source was copper (Cu) for the *in situ* XRD, emitting Cu $K\alpha_1$ and Cu $K\alpha_2$ radiation (ratio 2:1) with a wavelength of 1.5406 Å. The 2θ range was set between 15 and 100 degrees and a scan rate of 4° $min^{-1}$ was applied. $H_2$ and $C_2H_2$ were fed via MFCs (Bronkhorst EL-flow select). Applied CVD conditions were similar as in the hot-wall furnace system. Pumping employed a combined small turbomolecular pump/rotary value pump stage (turbomolecular: Oerlikon leybold vacuum turbovac T50, rotary).

***In situ* NAP XPS.** *In situ* NAP XPS experiments were performed at the CAT laboratory branches of the EMIL beamlines, UE48/PGM and CPMU17_EMIL, located at the synchrotron radiation facility BESSY II (Berlin, Germany). Applied CVD conditions were similar as in the hot-wall furnace system, albeit the NAP XPS reaction chamber has a base pressure of ~$10^{-8}$ mbar. The focus points of both beamlines meet in a dedicated NAP XPS analysis system based on a SPECS Phoibos 150 analyzer which covers the kinetic energy range up to 7 keV. The UHV system is described in detail elsewhere.[74,75] The sample temperature was measured through on surface clamped thermocouples, which however results in relatively large uncertainties for the Fe foils during graphene CVD (estimated ±100 °C). Thus stated sample temperatures were also corrected against *ex situ* growth results.[51] All XP spectra were recorded in normal photoemission geometry with a probing area of ≈ 60µm x 120µm corresponding to the profile of the incident x-ray beam. The overall spectral resolution of the NAP XPS system is about 0.4 – 0.5 eV at 10 eV pass energy. The binding energy (BE) was calibrated using the valence band onset of metallic Fe with a pronounced Fermi edge with an accuracy of around 0.05 eV. In order to get an overview of the sample, survey spectra were recorded using 1250 eV photon energy. Fe2p, O1s and C1s core levels were measured with 1250 eV, 1050 eV and 800 eV, respectively. Details on the XPS data analysis are given in the Supporting Information.

**ToF-SIMS.** ToF-SIMS was performed using a ToF-SIMS 5 instrument (IONTOF GmbH, Münster, Germany) equipped with a BiMn alloy liquid metal ion gun (LMIG), a dual-source column sputter gun (DSC) and an electron floodgun for charge compensation. A focused 25 keV Bi+ primary ion beam was employed to generate secondary ions, covering a mass range of m/z 1 to 230 (corresponding to a cycle time of 50 µs) in an analysis area of 100×100 µm. Measurements were carried out using sawtooth scanning with a pulse length of 13 ns in the high current bunched (HCBU) mode with a resolution of 128×128 pixels. This measurement mode provides high mass resolution (~11,000) and a pulsed LMIG current of ~2 nA. A 2 keV $Cs^+$ ion beam (300×300 µm, ~160 nA) was used for depth profiling since negative secondary ions like C- are detected more sensitively with $Cs^+$ bombardement.[76,77] To ensure charge compensation during depth profiling, an electron flood gun set at 21 V was activated in a non-interlaced cycle mode (5 s sputtering; 0.5 s pause). Samples were introduced into the instrument and allowed to equilibrate overnight, stabilizing the chamber pressure at $\sim 5\times10^{-9}$ mbar. The IONTOF ToF-SIMS instrument software, SurfaceLab 7 (version 7.1.130060), was used for data processing and mass calibration. The depth of the analysis craters was measured with a DekTakXT® profilometer (Bruker) to convert sputter time into depth.

**Nanoindentation.** Nanoindentation tests were carried out in an UMIS II nanoindentation system quipped with a Berkovich tip. Due to the rather thin samples, the load range was chosen to be 2 – 10 mN. Indents were made in steps of 0.5 mN. The recorded load-displacement curves were analysed using the procedure described by Oliver and Pharr.[78]

## Author Contributions

B.C.B. conceived the idea and supervised the work. B.F. conducted all experiments and data analysis. Author W.A. contributed to *in situ* XRD and D.M., J.R., M.N. and M.Ho. contributed to *in situ* XPS measurements. R.B., M.Hae., A.K.-G., B.R.C. and R.S. established the *in situ* XPS setup. A.K. and P.M. provided nanoindentation and F.F. and H.H. ToF-SIMS measurements, respectively. D.Z., C.D. and D.E. contributed to data interpretation. The manuscript was written by B.F. and B.C.B. with input from all authors.

## Acknowledgements

B.C.B., D.Z. and C.D. acknowledge funding from the Austrian Research Promotion Agency (FFG) under project 879844-HARD2D via the Production of the Future Programme of the Austrian Ministry of Climate Action, Environment, Energy, Mobility, Innovation and Technology (BMK). B.C.B. also acknowledges partial funding to the work from the European

Research Council (ERC) under project 101088366-HighEntropy2D. We thank the Helmholtz-Zentrum Berlin für Materialien und Energie for the allocation of synchrotron radiation beamtime at BESSY II (proposal 231-11784-ST) and Mihaela Gorgoi and Anna Efimenko for support. We acknowledge use of facilities at the X-ray Center (XRC) and Analytic Instrumentation Center (AIC) at TU Wien.

Supporting Information to:

# Realizing Scalable Chemical Vapour Deposition of Monolayer Graphene Films on Iron with Concurrent Surface Hardening by *in situ* Observations

Bernhard Fickl,[1] Werner Artner,[2] Daniel Matulka,[1,2] Jakob Rath,[1,3] Martin Nastran,[1] Markus Hofer,[1] Raoul Blume,[4] Michael Hävecker,[4,5] Alexander Kirnbauer,[6] Florian Fahrnberger,[7] Herbert Hutter,[7] Dengsong Zhang,[8,*] Paul H. Mayrhofer,[6] Axel Knop-Gericke,[4,5] Beatriz Roldan Cuenya,[5] Robert Schlögl,[4,5] Christian Dipolt,[9] Dominik Eder,[1,*] Bernhard C. Bayer[1,*]

[1]Institute of Materials Chemistry, Technische Universität Wien (TU Wien), Getreidemarkt 9, 1060 Vienna, Austria

[2]X-Ray Center, Technische Universität Wien (TU Wien), Vienna, Austria

[3]Analytical Instrumentation Center (AIC), Technische Universität Wien (TU Wien), Vienna, Austria

[4]Max-Planck-Institut für Chemische Energiekonversion, Postfach 101365, Mülheim an der Ruhr 45413, Germany

[5]Department of Inorganic Chemistry, Fritz-Haber-Institut der Max-Planck Gesellschaft, Faradayweg 4-6, 14195 Berlin, Germany

[6]Institute of Materials Science and Technology, Technische Universität Wien (TU Wien), Vienna, Austria

[7]Institute of Chemical Technologies and Analytics, Technische Universität Wien (TU Wien), Vienna, Austria

[8]International Joint Laboratory of Catalytic Chemistry, State Key Laboratory of Advanced Special Steel, Innovation Institute of Carbon Neutrality, Research Center of Nanoscience and Technology, Department of Chemistry, College of Sciences, Shanghai University, Shanghai 200444, China

[9]Rübig GmbH & Co KG, Schafwiesenstraße 56, 4600 Wels, Austria

*Corresponding Authors: bernhard.bayer-skoff@tuwien.ac.at (Bernhard C. Bayer), dominik.eder@tuwien.ac.at (Dominik Eder), dszhang@shu.edu.cn (Dengsong Zhang)

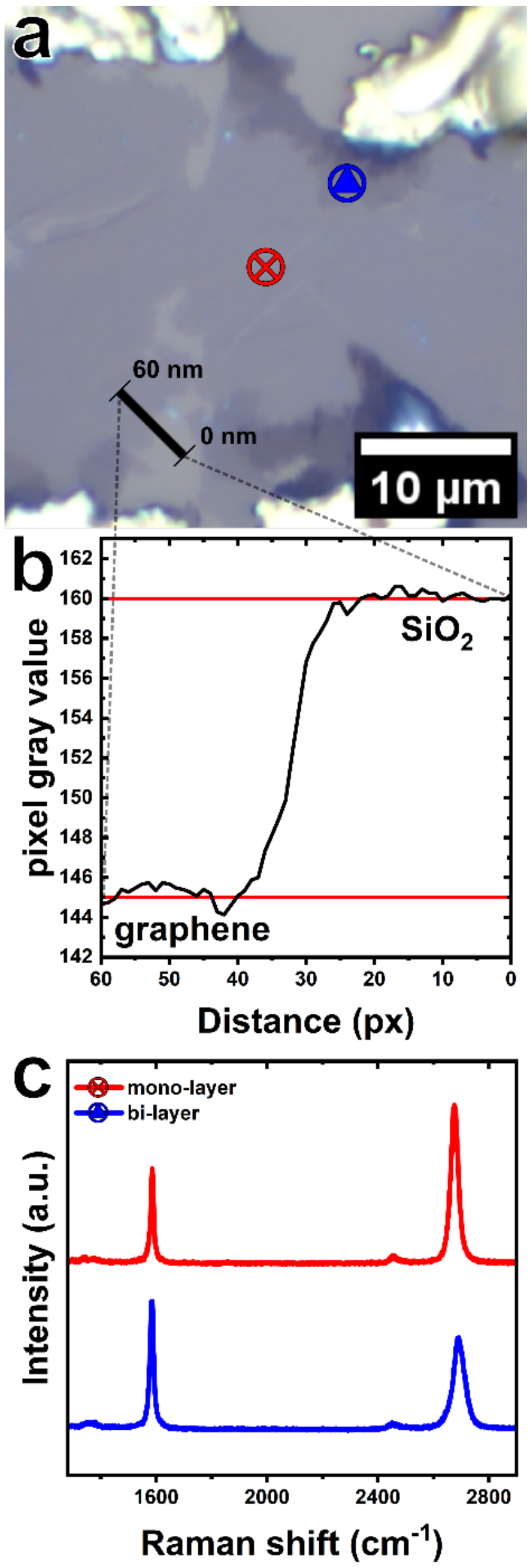

**Supporting Figure S1.** (a) Optical micrograph of CVD graphene from 750 °C/1 sccm $C_2H_2$ run from Figure 2 after polymer-assisted graphene transfer to 90 nm $SiO_2$-covered Si wafer. (b) Corresponding optical microscope pixel grey values averaged along the indicated black line in (a). (c) Corresponding point-localized Raman spectra with locations in (a) indicated by colour and symbol.

In Supporting Figure S1 we further characterize the optimized graphene films from the 750 °C and 1 sccm $C_2H_2$ runs in Figure 2. To this end, we transfer the films using a standard polymer-assisted transfer process[1] on 90 nm $SiO_2$-coated Si wafers. The optical images in Supporting Figure S1a together with the Raman point-localized spectra (red trace) in Supporting Figure S1c as well as the optical grey values in Supporting Figure S1b reaffirm that the deposited film is predominantly monolayer graphene of high crystalline quality (D/G <3%).[2] The monolayered nature of the graphene film is in particular confirmed by a 2D/G ratio of 2 and a 2D width of 28 $cm^{-1}$ that is readily fitted with a single Lorentzian (thus excluding formation of turbostratic multilayer graphene), see Supporting Figure S2.[3] Furthermore, also the grey value just in Supporting Figure S1b fully corresponds with the values expected for monolayer graphene.[4] In line with Figure 2, the remaining non-monolayer-graphene areas are isolated multilayer graphene islands (dark spots) and bare regions that we ascribe to prior Fe-oxide covered regions that did not nucleate graphene (plus losses of monolayer region from incomplete transfer).

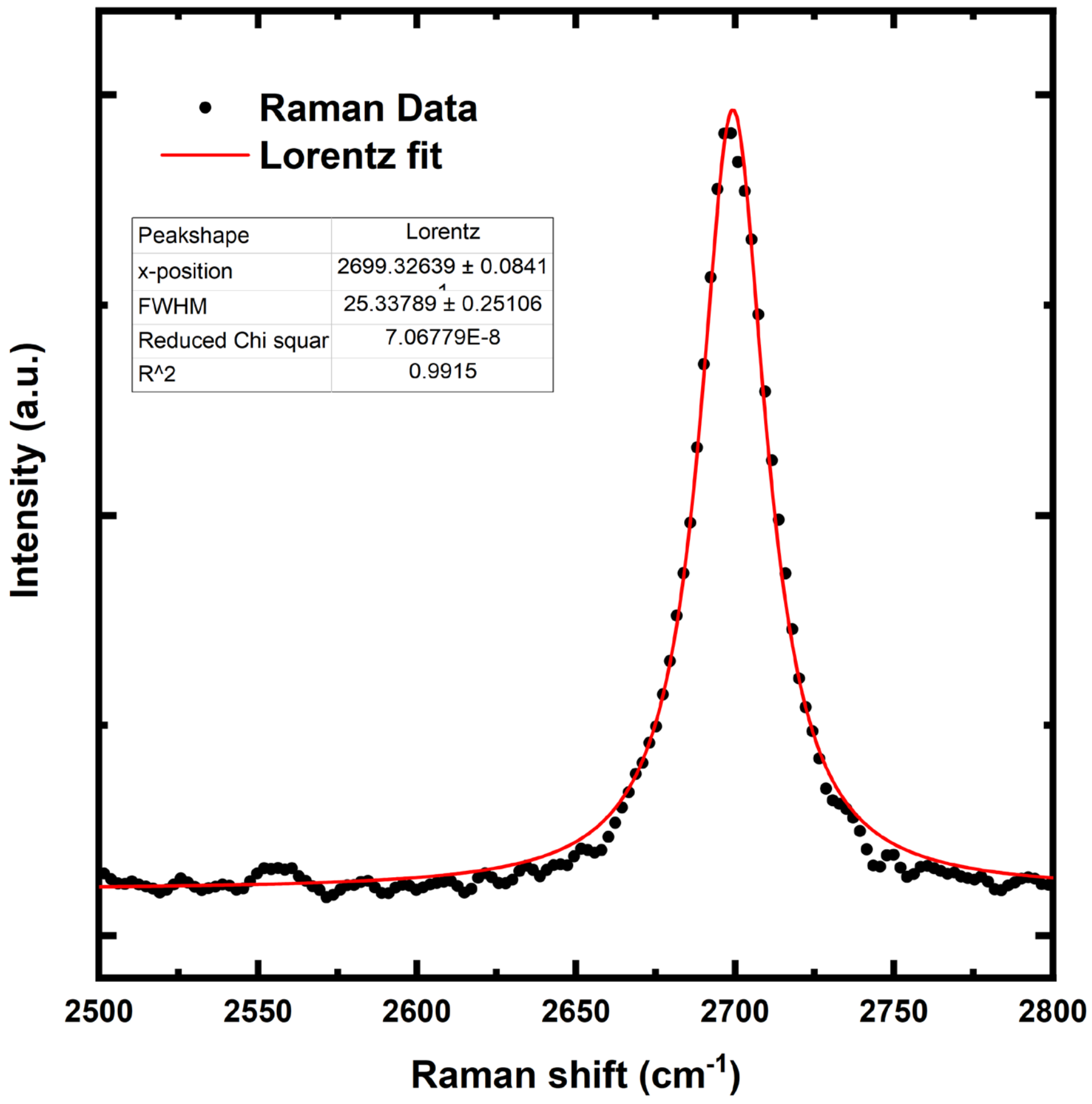

**Supporting Figure S2.** 2D peak of monolayer graphene film transferred from iron substrate onto $SiO_2$-coated Si wafer as in Supporting Figure S1 and fit with single Lorentzian (FWHM=25.3).[2]

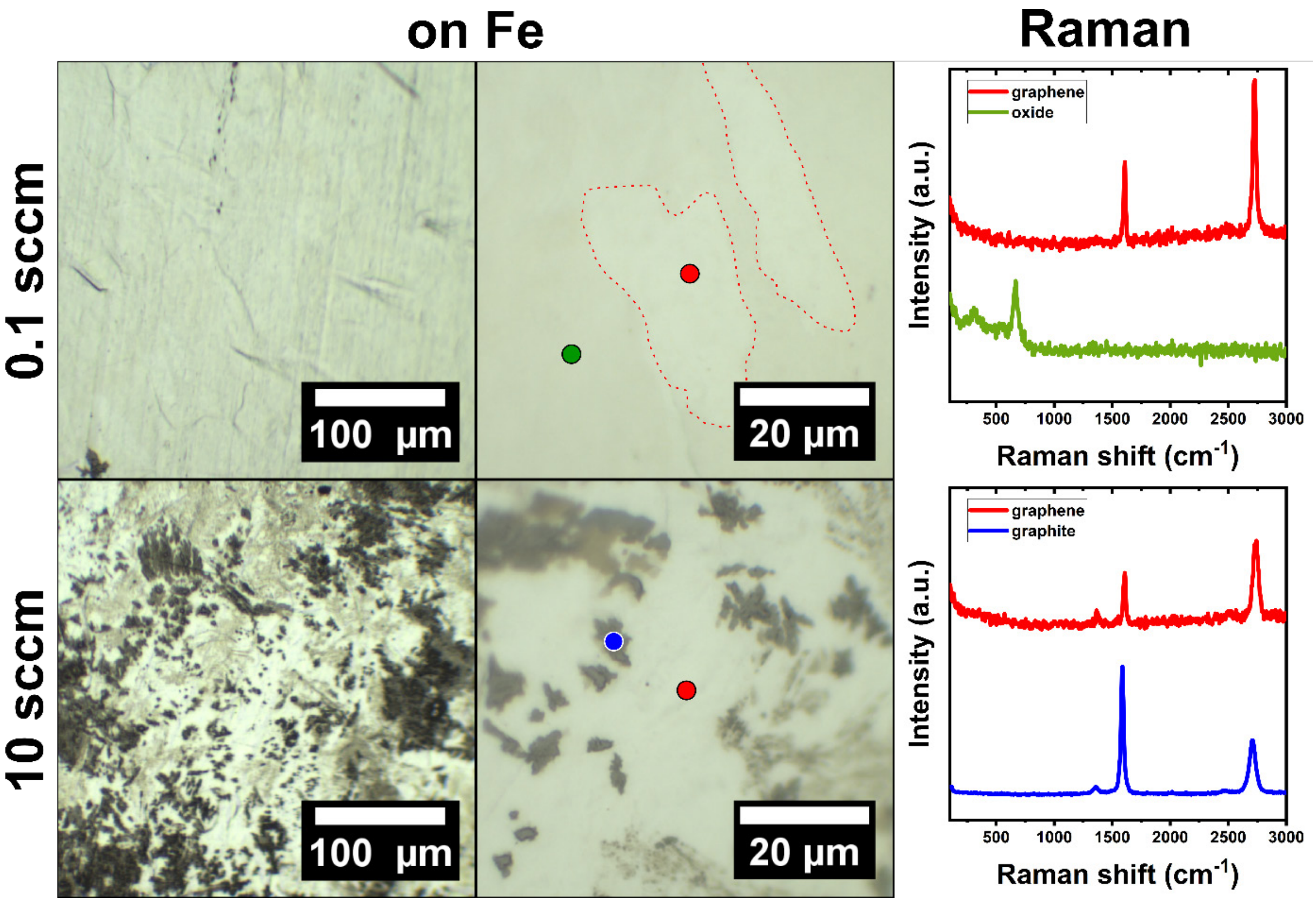

**Supporting Figure S3.** Optical micrographs (left and middle panels) at different magnifications and point localized Raman spectra (right panels) for 0.1 sccm (upper panels) and 10 sccm (lower panels) of $C_2H_2$ precursor flow at 750 °C CVD temperature. High quality monolayer graphene islands in upper middle panel are indicated by dashed red outlines. Upper right panel shows Raman spectra of monolayer graphene regions (red) and iron oxide regions (green). Lower right panel shows Raman spectra of graphitic multilayer regions (blue) and monolayer graphene regions (red).

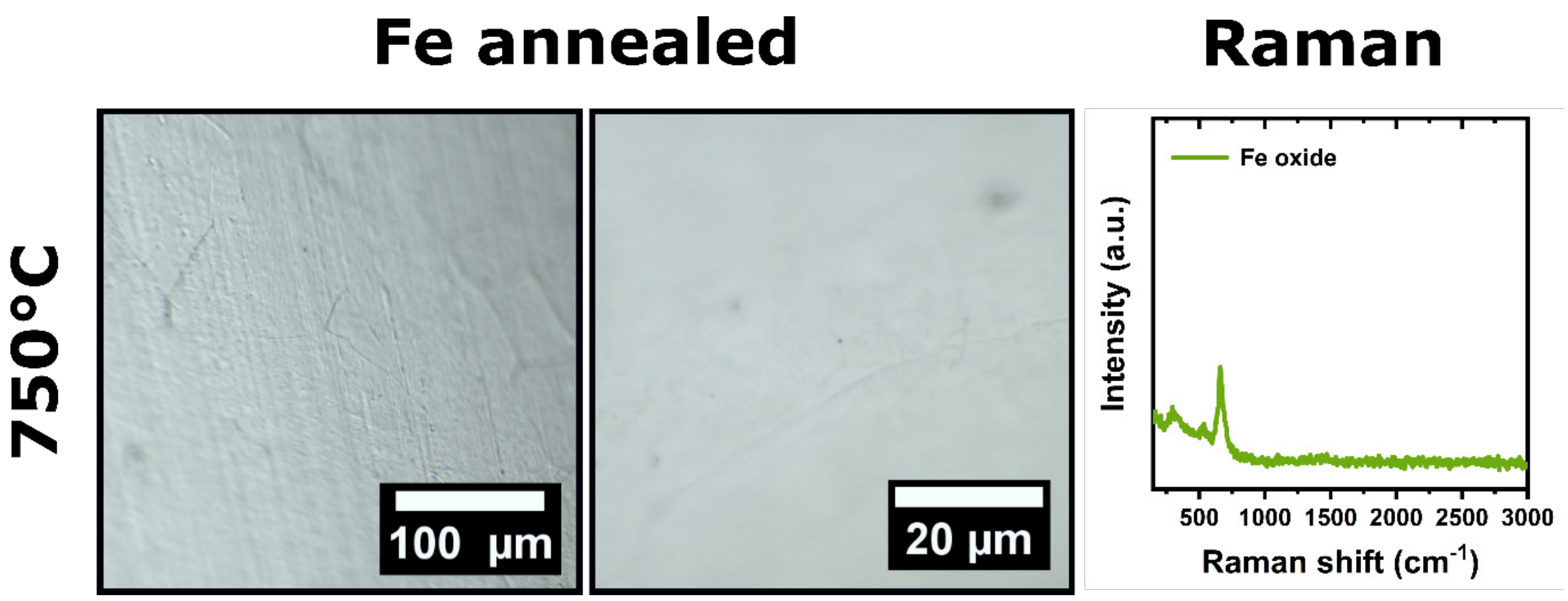

**Supporting Figure S4.** Left and middle panel show optical micrographs of $H_2$ annealed (750°C) Fe samples at different magnification respectively. Right panel shows a point localized Raman spectrum and iron-oxide signal (green trace), representative for the entire sample surface.

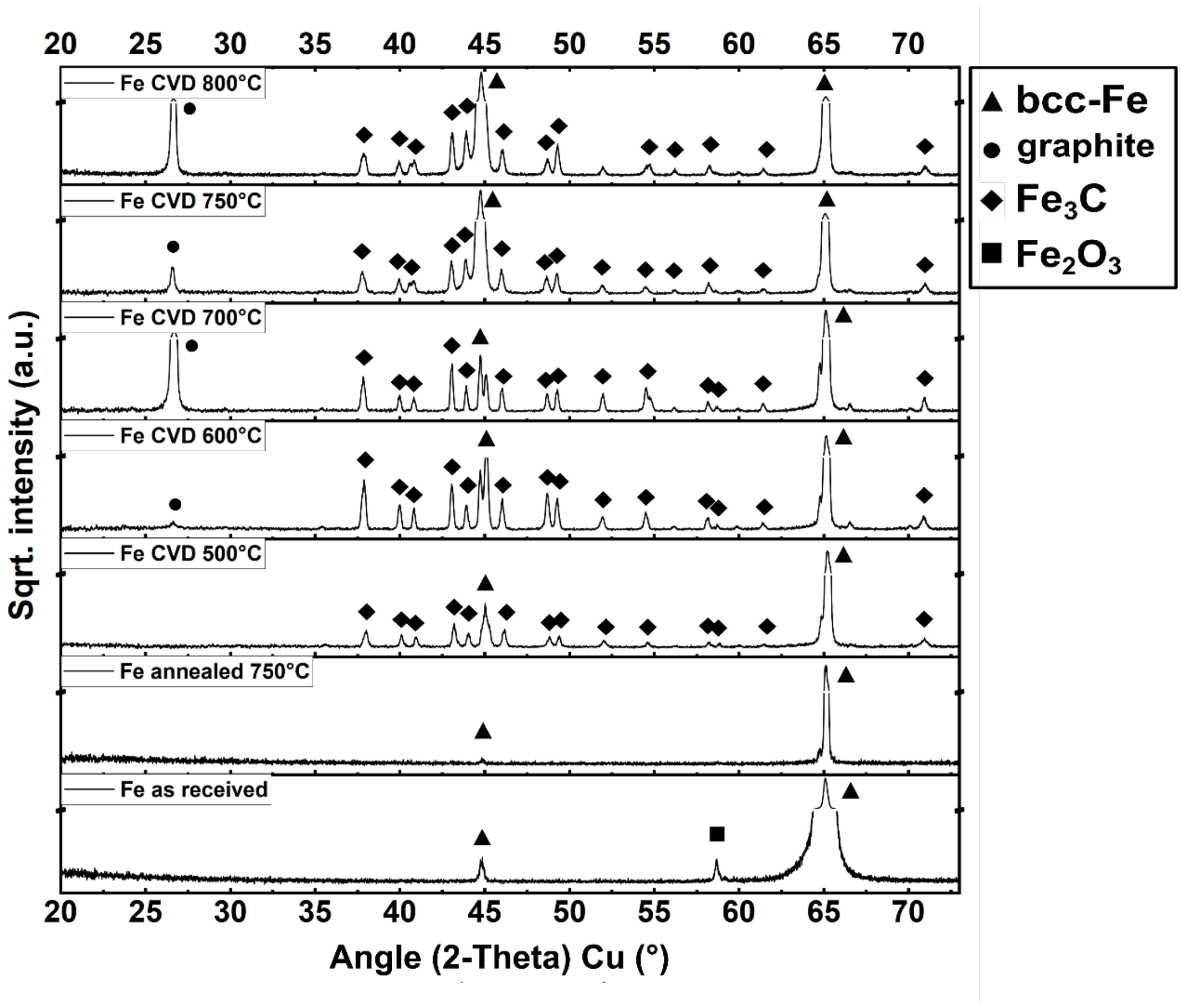

**Supporting Figure S5.** *Ex situ* XRD patterns of Fe catalysts supports after CVD conditions corresponding to Figure 2, as received Fe foil and after only $H_2$ treatment ("annealed at 750 °C"). Salient phases identified are indicated. (International Centre for Diffraction Data (ICDD), PDF-5+ database, powder diffraction file entry: bcc-Fe 04-015-8438; Carbon/graphite 04-016-0554; $Fe_3C$ 04-007-0422) Note that the intensity scale is plotted in square-root and the intensity scale interruption(s) for better visualization of minor $Fe_3C$ phase signal. Cr anode 2-Theta data was recalculated to Cu anode for better comparison.

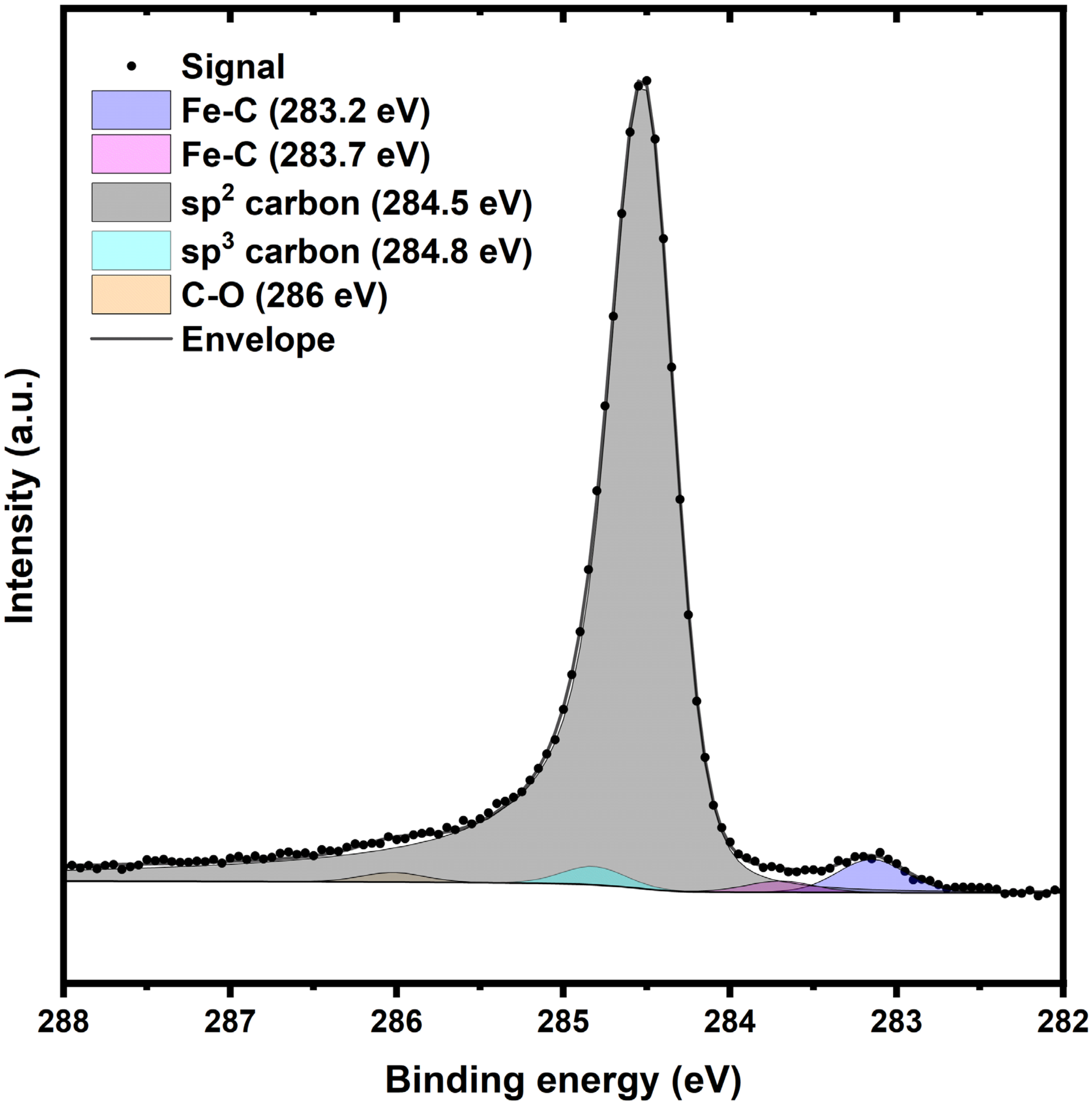

**Supporting Figure S6.** C1s spectrum during $C_2H_2$ exposure at 750 °C with peak components fit to experimental data, taken from the evolution in Figure 4a.

We identify in Supporting Figure S6 four primary components of our C1s peak related to graphene growth, plus one component considering the minimal contribution of C-O bound carbon. Firstly a peak at 283.2 eV which we ascribe to carbon bonded at iron surface sites based on previous work using Ni substrates.[5,6] On the same basis we identify a peak for carbon at interstitial Fe sites around 283.7 eV. We assign both components to carbon interacting with Fe and thus label both components "Fe-C". We include a $sp^2$-hybridized carbon peak at 284.5 eV and a $sp^3$-hybridized carbon peak at around 284.8 eV (consistent with remaining defects, graphene edges and grain boundaries in the graphene) together with a C-O component at around 286 eV in accordance with literature.[7]

Peak fitting was done in CasaXPS software[8] and the peak shapes are given with the software command abbreviations. GL(30) denoting a Gaussian/Lorentzian product with 30% Lorentzian contribution. LF denotes an asymmetric Lorentzian lineshape with a tail dampening parameter.[9,10] Full-width-at-half-max (FWHM) of $sp^3$ peak was constrained to FWHM of adventitious carbon peak, measured initially on bare Fe and the C-O component was constrained to follow the $sp^3$ component's FWHM. Fe-C components were also constrained to same FWHM based on FWHM of peak at 283.2 eV. General peakshapes were based on work from Gengerbach et al..[7]

| Peak designation | Peak position | Peak shape | FWHM |
|---|---|---|---|
| Fe-C | 283.2 eV ± 0.1 eV | GL(30) | 0.48 eV |
| Fe-C | 283.7 eV ± 0.1 eV | GL(30) | 0.48 eV |
| $sp^2$ | 284.5 eV ± 0.1 eV | LF(0.65,1.1,500,180,3) | 0.45 eV |
| $sp^3$ | 284.8 eV ± 0.1 eV | GL(30) | 0.69 eV |
| C-O | 286 eV ± 0.1 eV | GL(30) | 0.69 eV |

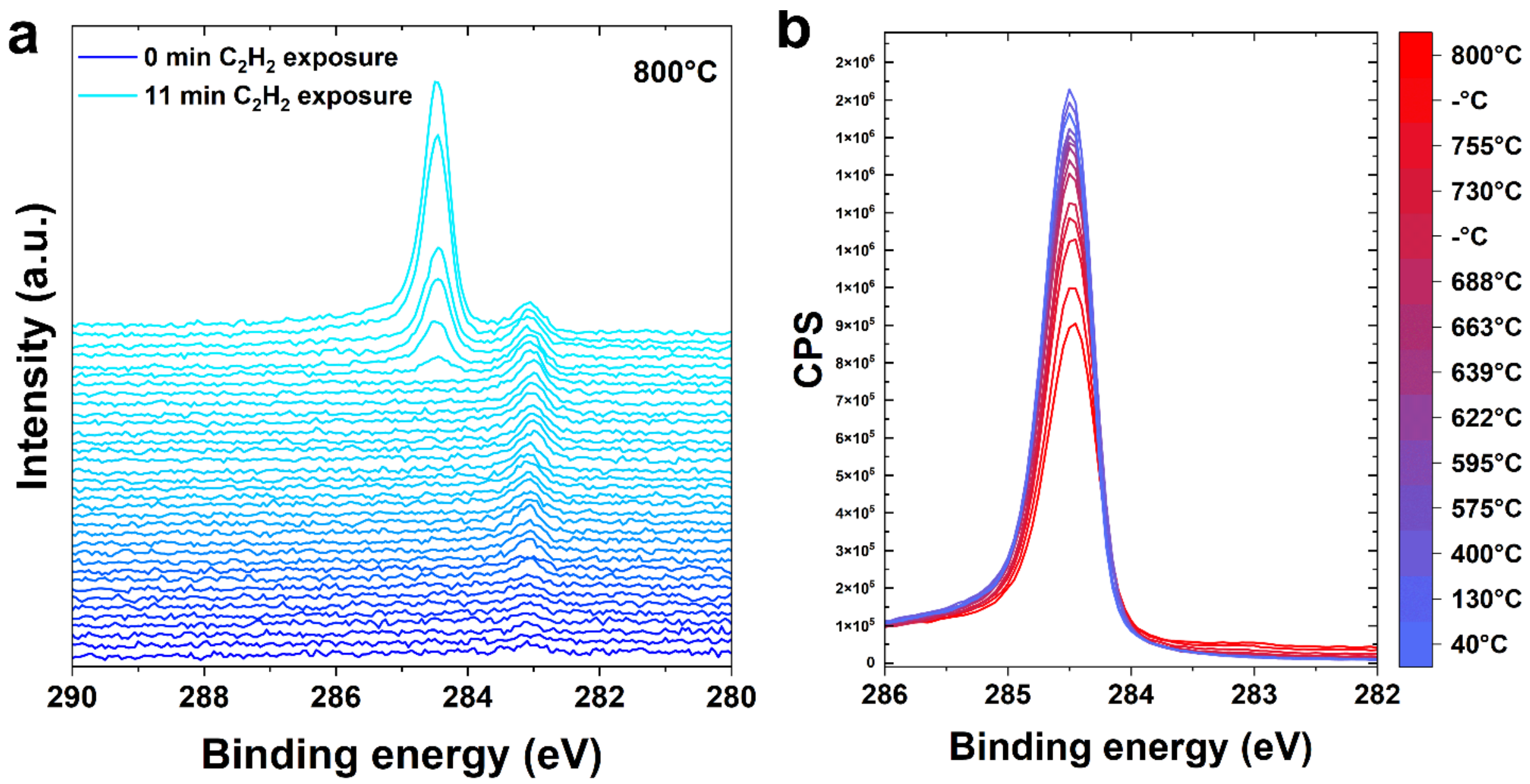

**Supporting Figure S7.** (a) Time resolved in-situ XPS C1s spectra during $C_2H_2$ exposure at 800°C showing an incubation of ~10 minutes from hydrocarbon exposure start to start of isothermal surface carbon growth. Time increasing from 0 s to 11 min of exposure from bottom to top. (b) C1s time resolved spectra during substrate cooling in $H_2$ atmosphere from 800 °C after $C_2H_2$ shut off show significant surface carbon intensity increase over time, reaching a plateau at around 400°C.

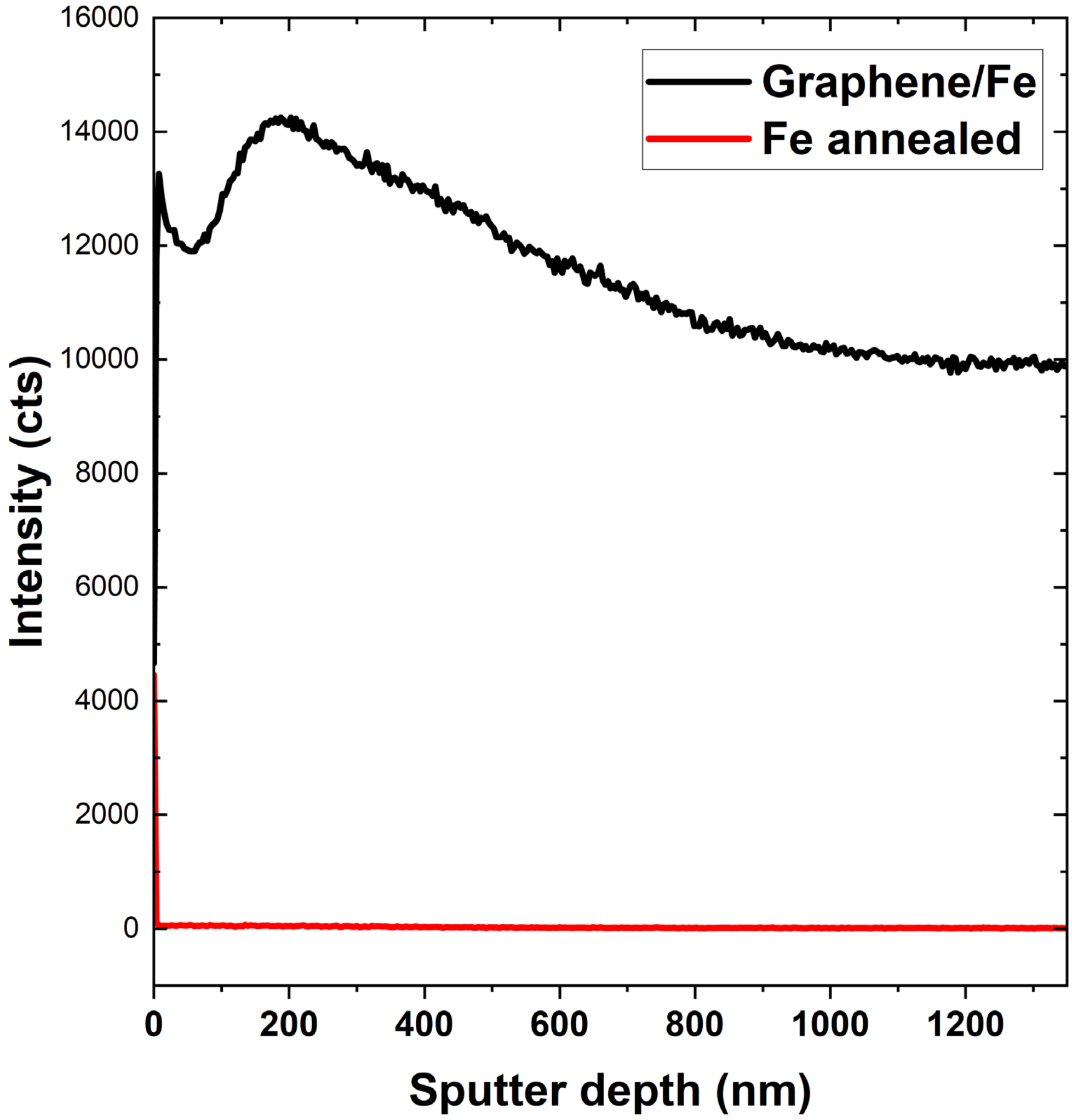

**Supporting Figure S8.** ToF-SIMS carbon anion $C^-$ depth profile of $H_2$ annealed (without $C_2H_2$) reference Fe foil (red trace) and Fe foil after optimized CVD (black trace). The $H_2$ annealed Fe sample shows no significant carbon content. The CVD iron foil shows a large carbon signal and depth dependent decrease indicating a carbon diffusion gradient.

## Supporting References